\documentclass
[aps,prc,twocolumn,amsmath,amssymb,floatfix]
{revtex4-1}

\usepackage{CJK}
\usepackage{mathptmx}
\usepackage[dvips]{graphicx}
\usepackage{dcolumn}
\usepackage{bm}
\usepackage{xcolor}
\usepackage{multirow}

\def\be{\begin{equation}}
\def\ee{\end{equation}}
\def\bal#1\eal{\begin{align}#1\end{align}}
\def\rv{\bm{r}}
\def\Jv{\bm{J}}
\def\Wv{\bm{W}}
\def\eps{\varepsilon}
\def\la{\Lambda}
\def\ron{\rho_N}
\def\rol{\rho_\Lambda}
\def\fm3{\,\text{fm}^{-3}}
\def\bav{\langle B_\la \rangle}

\begin{document}

\begin{CJK}{UTF8}{gbsn}

\title{
Effects of $\bm\la$ hyperons on the deformations of even-even nuclei}

\author{C. F. Chen (陈超锋)}
\author{Q. B. Chen (陈启博)}
\author{Xian-Rong Zhou (周先荣)} \email{xrzhou@phy.ecnu.edu.cn}
\author{Yi-Yuan Cheng (程奕源)}
\affiliation{Department of Physics, East China Normal University,
	Shanghai 200241, China}

\author{Ji-Wei Cui (崔继伟)}
\affiliation{School of Physics and Optoelectronic Engineering,
Xidian University, Xi'an 710071, China}

\author{H.-J. Schulze}
\affiliation{
INFN Sezione di Catania, Dipartimento di Fisica,
Universit\'a di Catania, Via Santa Sofia 64, 95123 Catania, Italy}

\date{\today}

\begin{abstract}
The deformations of multi-$\la$ hypernuclei corresponding to even-even
core nuclei ranging from $^8$Be to $^{40}$Ca with 2, 4, 6, and 8 hyperons
are studied in the framework of the deformed Skyrme-Hartree-Fock approach.
It is found that the deformations are reduced when adding 2 or 8 $\la$ hyperons,
but enhanced when adding 4 or 6 $\la$ hyperons.
These differences are attributed to the fact that $\la$ hyperons
are filled gradually into the three deformed $p$ orbits,
of which the [110]1/2$^-$ orbit is prolately deformed
and the degenerate [101]1/2$^-$ and [101]3/2$^-$ orbits are oblately deformed.
\end{abstract}

\maketitle

\end{CJK}

\section{Introduction}

Since the first discovery of a hypernucleus in cosmic rays \cite{danysz1953},
the study of hypernuclei has become one of the most interesting topics in
nuclear physics from both experimental and theoretical sides
\cite{hashimoto2006,Hiyama2010PTPS,Hiyama2010PTPS_v1,Tamura2012PTEP,
Feliciello2015RPP,gal2016,hiyama2018}.
In particular, $\la$ hypernuclei have been studied by many experiments
and related theoretical analyses.
Due to the limitation of current experimental conditions,
the experimental data of hypernuclei are mainly for single-$\la$ hypernuclei
\cite{hashimoto2006,Hiyama2010PTPS,bertini1979,davis1986,pile1991,hasegawa1996,
Cusanno2009,agnello2011,Feliciello2015RPP,gal2016},
with only a few data for double-$\la$ hypernuclei
\cite{danysz1963,prowse1966,franklin1995,takahashi2001,Hiyama2010PTPS_v1,
ahn2013,Feliciello2015RPP,yoshida2016,hiyama2018,ekawa2019}.
Of particular interest is the fact that the
addition of $\la$ hyperons can lead to the appearance of the
so-called impurity effect \cite{hashimoto2006,gal2016},
since the hyperon(s) can enter deeply into the center of a hypernucleus
regardless of the restriction of the Pauli exclusion principle,
and thus can be used as a good probe to study the nuclear environment.
The impurity effects of single-$\la$ hypernuclei have been investigated
extensively in the past decades,
such as the shrinkage effect \cite{motoba1985,hiyama1999,tanimura2019},
the modification of the drip lines \cite{vretenar1998,Zhou08},
and the modification of the deformation
\cite{win2008,schulze2010,win2011,isaka2011,lu2011,isaka2012,cui2017}.

Since the experimental discovery of double-$\la$ hypernuclei
\cite{danysz1963,prowse1966,franklin1995,takahashi2001,ahn2013,
yoshida2016,ekawa2019},
several kinds of nuclear models have been extended to study the $\la\la$
hypernuclei sector.
For example, the microscopic cluster model was used
to study the $\la$ binding energies of light hypernuclei and
reproduced well the observation of the ground state of
$^{11}_{\la\la}$Be \cite{hiyama2002,hiyama2010}.
The Faddeev calculations with the Nijmegen soft-core potential NSC97
described well the binding energies of light $\la\la$ hypernuclei \cite{gal2005}.
The shell-model calculation showed how the $\la N$ spin-dependent
interaction terms influence the $\la\la$ hypernuclei across the
nuclear $p$ shell \cite{gal2011}.
The beyond-mean-field approach was
applied to study the evolution of nuclear deformation in $\la\la$ hypernuclei
and the hyperon impurity effect in hypernuclei with shape coexistence
\cite{wu2017,mei2018}.
Extensive research of binding energies and deformation effects has been
carried out by the self-consistent mean-field model on the shape of hypernuclei
\cite{cugnon2000,vida2001,Zhou08,win2008,win2011,schulze2013,zhou2016,sun2016}.

Very recently, the impurity effects of multi $\la$ hyperons on the
deformations in the hyperisotope chains
$^{8+n}_{\ \ n\la}$Be ($n=2,4$),
$^{20+n}_{\ \ \ \ n\la}$Ne ($n=2$,4,8), and
$^{28+n}_{\ \ \ \ n\la}$Si ($n=2$,4,8)
have been studied using the relativistic mean field (RMF) theory
in Ref.~\cite{tanimura2019}.
It was pointed out that in the Ne hyperisotopes,
the deformation is slightly reduced by the additional $\la$ hyperons,
whereas it is significantly reduced or even disappears in the Si hyperisotopes.
Studies on multi-$\la$ hyperisotopes have theoretical significance,
although the corresponding experiments are currently unfeasible.
First of all, the impurity effects in multi-$\la$ hyperisotopes
are evidently stronger than those of single-$\la$ ones.
Moreover, a multi-$\la$ system can provide important information
on the $\la\la$ interaction,
and the effects of the core nucleus on the $\la$ hyperons can be studied.

The aforementioned work in Ref.~\cite{tanimura2019} studied only a few nuclei
and did not address the impurity effects caused by 6 $\la$ hyperons.
Therefore, further studies on the impurity effect of
$n=2, 4, 6, 8\;\la$ hyperons on the properties of even-even
nuclei ranging from $^8$Be to $^{40}$Ca are carried out in this work.
In contrast to the RMF model
adopted in Ref.~\cite{tanimura2019},
we will employ the deformed Skyrme-Hartree-Fock (SHF) approach
\cite{vautherin1972,vautherin1973,bender2003},
which is one of the widely used models for hypernuclei
\cite{zhou2007,Zhou08,zhou2016}.

\section{Formalism}

In the framework of the SHF approach,
the energy of a hypernucleus is given by an energy-density functional,
\be
 E = \int d^3\rv\;\eps(\rv)
\ , \quad
 \eps = \eps_{NN} + \eps_{\la N} + \eps_{\la \la} \:,
\label{one}
\ee
where $\eps_{NN}$, $\eps_{\la N}$, and $\eps_{\la\la}$
account for the nucleon-nucleon interaction,
the hyperon-nucleon interaction,
and the hyperon-hyperon interaction, respectively.
The energy-density functional
depends on the one-body densities $\rho_q$,
kinetic densities $\tau_q$,
and spin-orbit currents $\Jv_q$,
\be
  \Big[ \rho_q,\; \tau_q,\; \Jv_q \Big] =
  \sum_{i=1}^{N_q} {n_q^i} \Big[
  |\phi_q^i|^2 ,\;
  |\nabla\phi_q^i|^2 ,\;
  {\phi_q^i}^* (\nabla \phi_q^i \times \bm{\sigma})/i
 \Big] \:,
\label{e:rho}
\ee
where $\phi_q^i$ ($i=1,N_q$) are the self-consistently calculated
single-particle (s.p.) wave functions
of the $N_q$ occupied states for the species $q=n,p,\la$
in a hypernucleus.
They satisfy the Schr\"odinger equation,
obtained by the minimization of the total energy functional (\ref{one})
according to the variational principle, as
\be
 \bigg[ \nabla \cdot {1\over2m_q^*(\rv)}\nabla - V_q(\rv)
 + i \Wv_q(\rv) \cdot \left(\nabla\times\bm{\sigma}\right)
 \bigg] \phi_q^i(\rv)
 = e_q^i \,\phi_q^i(\rv)
\ee
in which $\Wv_q(\rv)$ is the spin-orbit interaction part for the
nucleons as given in Refs.~\cite{vautherin1972,bender1999}.
The central mean fields $V_q(\rv)$,
corrected by the effective-mass terms
following the procedure described in
\cite{cugnon2000,schulze2013,margueron2017}
are
\bal
 V_N =&
 V_N^{\text{SHF}} + \frac{\partial\eps_{N\la}}{\partial\ron}
\notag\\
 &+ \frac{\partial}{\partial\ron} \left(\frac{m_\la}{m_\la^*(\ron)}\right)
 \left( \frac{\tau_\la}{2m_\la}
 - \frac{3}{5}\frac{\rol(3\pi^2\rol)^{2/3}}{2m_\la} \right) \:,
\\
 V_\la =&
 \frac{\partial (\eps_{N\la}+\eps_{\la\la})}{\partial\rol}
 - \left( \frac{m_\la}{m_\la^*(\ron)} - 1 \right)
 \frac{(3\pi^2\rol)^{2/3}}{2m_\la} \:.
\label{e:Vla}
\eal

\begin{table}[t]
\caption{
Parameters of the NSC97f+EmpC functionals \cite{vida2001,margueron2017}
of energy density and $\la$ effective mass,
Eqs.~(\ref{e:eln},\ref{e:ell},\ref{e:mfit}),
used in this work.}
\label{t:para}
\begin{ruledtabular}
\begin{tabular}{c c c c c c c c c c }
 $\eps_1$ & $\eps_2$ & $\eps_3$ & $\eps_4$ & $\eps_5$ & $\eps_6$ & $\eps_7$ &
 $\mu_1$ & $\mu_2$ & $\mu_3$
\\
 384 & 1473 & 1933 & 635 & 1829 & 4100 & 33.25 & 0.93 & 2.19 & 3.89 \\
\end{tabular}
\end{ruledtabular}
\end{table}

For the nucleonic part $\eps_{NN}$,
we use the Skyrme force SLy5 \cite{chabanat1998,bender2003},
which has been fitted in a wide nuclear region.
The energy-density contributions
$\eps_{N\la}$ \cite{vida2001,schulze2013} and
$\eps_{\la\la}$ \cite{margueron2017}
are parameterized as
($\rho$ given in units of fm$^{-3}$,
 $\eps$ in $\rm MeV\,fm^{-3}$):
\bal
 \eps_{N\la}(\ron,\rol) =& -( \eps_1 - \eps_2\ron + \eps_3\ron^2) \ron\rol
\nonumber\\
              & +( \eps_4 - \eps_5\ron + \eps_6\ron^2) \ron\rol^{5/3} \:,
\label{e:eln}
\\
 \eps_{\la\la}(\rol) =& -\eps_7 \rol^2 \Theta(N_\la>1)\:,
\label{e:ell}
\eal
together with
\bal
  {m_\la^*\over m_\la}(\ron) &\approx
  \mu_1 - \mu_2\ron + \mu_3\ron^2   \:.
\label{e:mfit}
\eal
The parameters $\eps_1,\ldots,\eps_6$ in Eq.~(\ref{e:eln})
and the $\la$ effective-mass parameters $\mu_i$ were determined in
Brueckner-Hartree-Fock calculations
of hypernuclear bulk matter
with the Nijmegen potential NSC97f \cite{vida2001,schulze2013},
while the empirical expression involving the parameter $\eps_7$
in Eq.~(\ref{e:ell}) 
has been proposed by fitting the bond energy of $^{\ \ 6}_{\la\la}$He
in Ref.~\cite{margueron2017}.
All parameters are listed in Table~\ref{t:para}.
This procedure gives a good description of the binding energies of single- and
double-$\la$ hypernuclei \cite{cugnon2000,schulze2013,margueron2017}.

The occupation probabilities $n^i_q$ (for nucleons only)
in Eq.~(\ref{e:rho}) are calculated by taking into account
pairing interactions within a BCS approximation.
In this work,
the pairing interaction is taken as a density-dependent
$\delta$ interaction \cite{tajima1993},
\be
 V_{q}(\rv_1,\rv_2) =
 V_0 \left[ 1 - \frac{\ron((\rv_1+\rv_2)/2)}{0.16\;\text{fm}^{-3}} \right]
 \delta(\rv_1-\rv_2) \:.
\ee
For the $p$-shell nuclei and their corresponding hypernuclei,
the strength of the pairing force is set to
$V_0=-410$~MeVfm$^3$ for both neutrons and protons,
which gives reasonable binding energies for $^{12}$C and
$^{13}_{\ \la}$C \cite{win2011,sagawa2004}.
For the heavier (hyper)nuclei,
$V_0$ is taken as $-999\;$MeVfm$^3$ for neutrons and
$-1146\;$MeVfm$^3$ for protons as in Ref.~\cite{bender2000}.

In this work, we focus mainly on the impurity effects of multi $\la$
hyperons on the deformation of nuclei.
The deformed SHF Schr\"odinger equation is solved in cylindrical coordinates
$(r,z)$ under the assumption of axial symmetry of the mean field
\cite{vautherin1973,bender2003}.
The optimal quadrupole deformation parameter
\be
 \beta_2 = \sqrt{\frac{\pi}{5}}
 \frac{\langle 2z^2-r^2 \rangle}{\langle z^2+r^2 \rangle}
\ee
is determined by minimizing the energy-density functional.

\section{Results}

\begin{figure}[t]
\includegraphics[width=85mm]{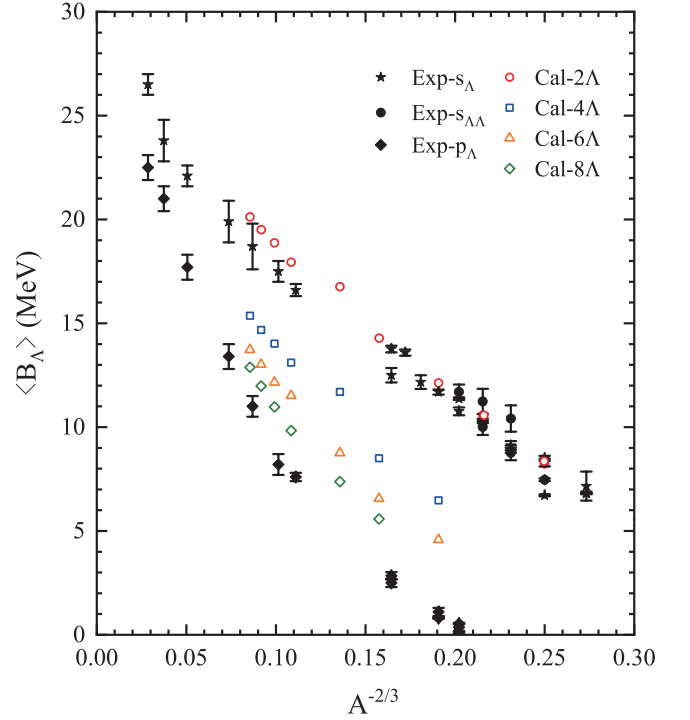}
\vskip-2mm
\caption{
Average $\la$ binding energies in multi-$\la$ hypernuclei
$\bav \equiv B_{n\la}/n$
as functions of $A^{-2/3}$ calculated by SHF
in comparison with the experimental data
of single-$s_\la$ hypernuclei,
double-$s_{\la\la}$ hypernuclei,
and single-$p_\la$ hypernuclei.
The experimental data are taken from
Ref.~\cite{gal2016} and references therein.}
\label{f:binding}
\end{figure}

\begin{figure*}[t]
\includegraphics[width=180mm]{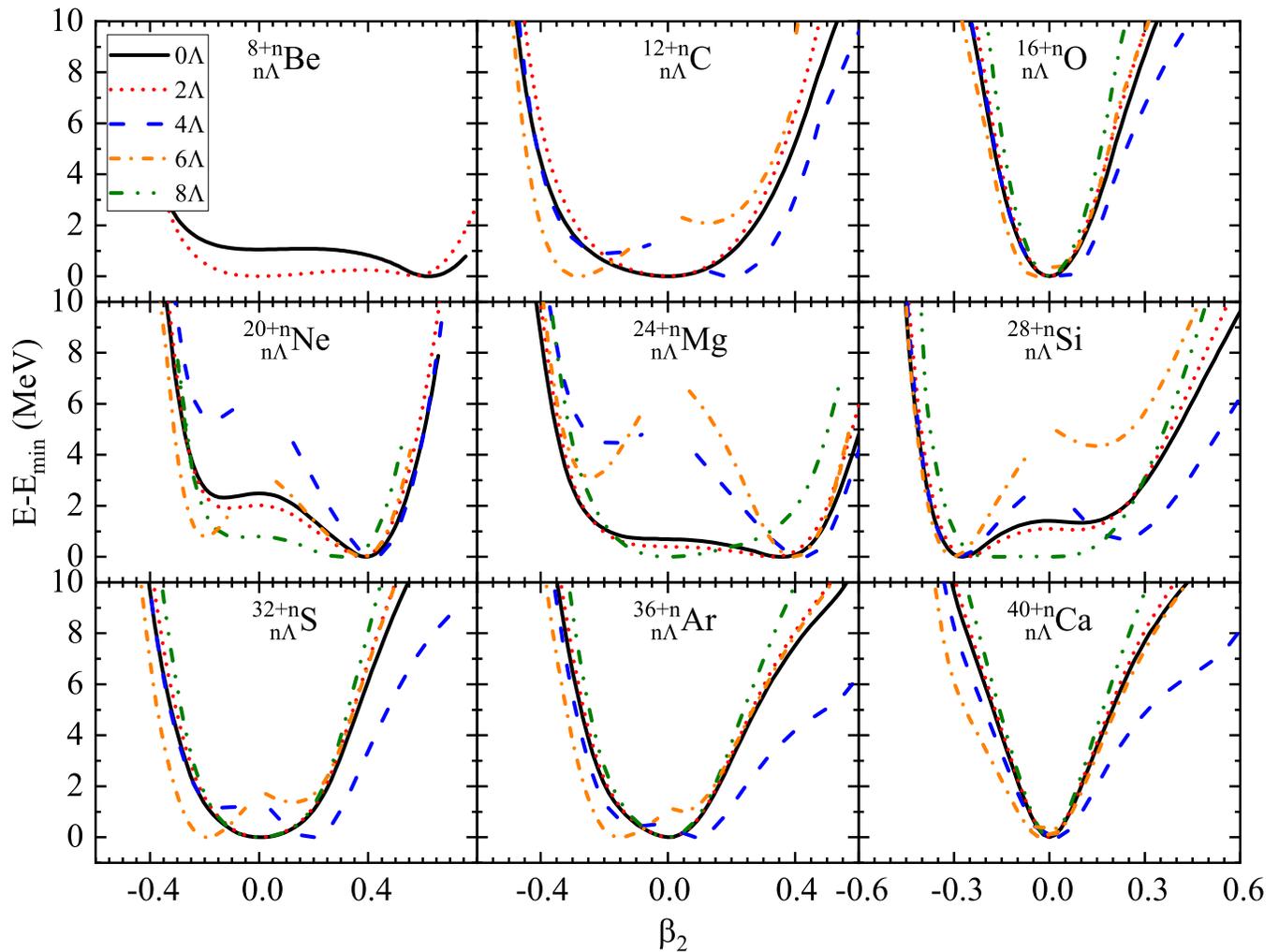}
\vskip-2mm
\caption{
Potential energy surfaces as functions of quadrupole deformation $\beta_2$
calculated by the self-consistent deformed SHF method
for even-even nuclei ranging from $^8$Be to $^{40}$Ca
and their corresponding multi-$\la$ ($n=2$, 4, 6, 8) hypernuclei.
Energies are normalized with respect to the binding energy
of the absolute minimum for a given isotope.
Positive (negative) values of $\beta_2$ correspond to
prolate (oblate) deformation.}
\label{f:eng}
\end{figure*}

Due to the lack of experimental data of multi-$\la$ hypernuclei,
we compare in Fig.~\ref{f:binding}
the average $\la$ binding energy of multi-hyperon hypernuclei,
$\bav \equiv B_{n\la}/n$,
with that of experimental single-$s_\la$ hypernuclei,
double-$s_{\la\la}$ hypernuclei,
and single-$p_\la$ hypernuclei.
Both theoretical and experimental results
show that $\bav$ decreases with $A^{-2/3}$.
Due to the weak $\la\la$ interaction,
the $\bav$ values of double-$s_{\la\la}$ hypernuclei are very close to those of
single-$s_\la$ hypernuclei,
For a given isotope, $\bav$ decreases with increasing hyperon number,
since the higher $\la$ s.p.~orbits are being filled.
As a consequence, the $\bav$ of 8-$\la$ hypernuclei are
close to those of the single-$p_\la$ hypernuclei.
These comparisons are rather qualitative,
and experimental binding energies of multi-$\la$ hypernuclei
are necessary to do a strict evaluation
of the current theoretical calculations.
Nevertheless,
as the energies of single- and double-$\la$ hypernuclei
are reasonably well reproduced,
we continue the analysis for other quantities based on the current model.

The impurity effect of additional hyperons in single-$\la$ or
double-$\la$ hypernuclei
is usually reflected by the shape shrinkage or deformation reduction
of the nuclear core.
To study the impurity effect of multi-$\la$ systems,
we show in Fig.~\ref{f:eng} the calculated potential energy surfaces as
functions of the quadrupole deformation $\beta_2$
for even-even nuclei ranging from $^8$Be to $^{40}$Ca
and their corresponding multi-$\la$ ($n_\la = 2,4,6,8$) hypernuclei.
All energies are normalized with respect to the binding energy
of the absolute minimum for a given isotope.
Apart from $^{12}$C, $^{32}$S, $^{36}$Ar,
and the doubly magic nuclei
$^{16+n}_{\ \ \ n\la}$O and
$^{40+n}_{\ \ \ n\la}$Ca ($n=0,2,4,6,8$),
all other (hyper)nuclei are well deformed.
$^{12+n}_{\ \ \ n\la}$C ($n=2,4,6$),
$^{28+n}_{\ \ \ n\la}$Si ($n=0,2,4,6,8$), and
$^{42}_{6\la}$Ar are oblately deformed,
while the others are prolately deformed.

One observes that the impurity effects become stronger with more hyperons
involved, but the dependence is not regular:
For $2\la$ and $8\la$ hypernuclei,
the impurity effect gives similar results
of deformation reduction as in the case of single-$\la$ hypernuclei.
This observation is the same as that obtained by the
RMF model in Ref.~\cite{tanimura2019}.
However, for $4\la$ and $6\la$ hypernuclei,
the opposite impurity effects can be seen, namely
the deformations of the hypernuclei become larger than those of the core nuclei.
The energy differences between the prolate and oblate local minima
in Ne, Mg, and Si isotopes are smaller than 2 MeV,
which characterize them as typical nuclei
with shape-coexistence phenomenon \cite{heyde2011}.
With the addition of hyperons,
not only these nuclei retain their shape coexistence,
but also other hypernuclei,
such as C, O, S, and Ar,
can develop the shape-coexistence phenomenon.

\begin{figure}[t]
\includegraphics[width=90mm]{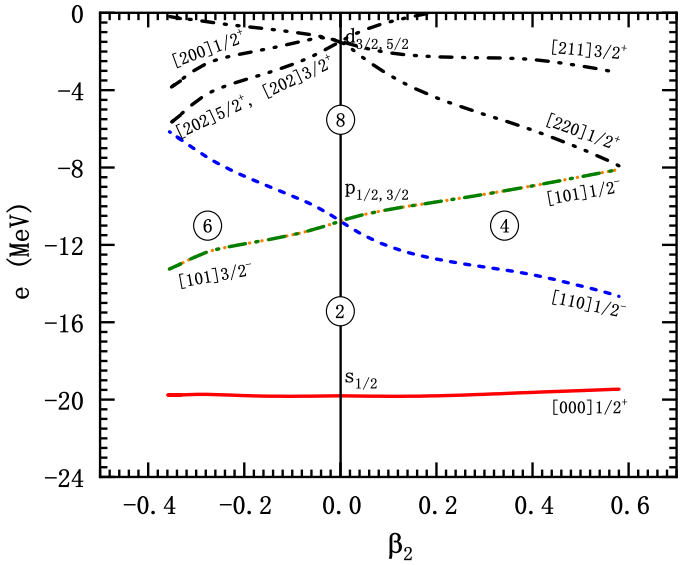}
\vskip-3mm
\caption{
The calculated $\la$ hyperon $s,p,d$ s.-p.~energy levels
as function of quadrupole deformation $\beta_2$
in $^{48}_{8\la}$Ca.}
\label{f:nil}
\end{figure}

\begin{figure}[t]
\includegraphics[width=85mm]{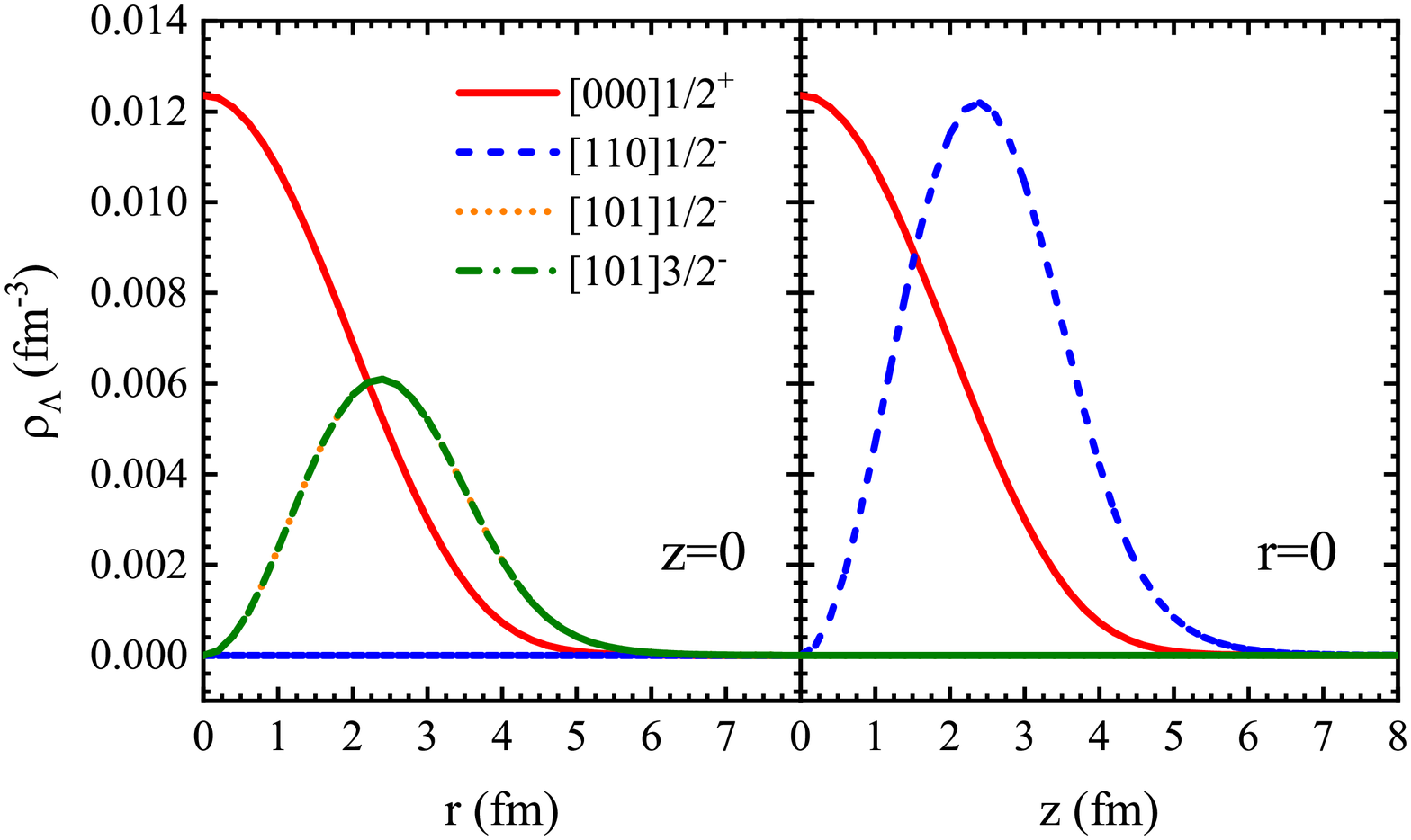}
\vskip-3mm
\caption{
Density distributions
for the occupied $s$ and $p$ s.-p.~orbits
of $\la$ hyperons in $^{48}_{8\la}$Ca at $\beta_2=0$
as functions of $r$ ($z=0$) and $z$ ($r=0$).
The $z$ axis is the symmetry axis.}
\label{f:orbit}
\end{figure}

In order to achieve a microscopic understanding of the behavior of $\bav$
in Fig.~\ref{f:binding}
and the impurity effects of multi $\la$'s on the deformation in Fig.~\ref{f:eng},
we take $^{48}_{8\la}$Ca as example,
and show in Fig.~\ref{f:nil}
the s.p.~energies of $\la$ hyperons as a function of $\beta_2$,
and in Fig.~\ref{f:orbit}
the density distributions at $\beta_2=0$
as functions of $r$ ($z=0$) and $z$ ($r=0$)
for the occupied $s$ and $p$ $\la$ s.p.~orbits.
Note that the $z$ axis is the symmetry axis.

Fig.~\ref{f:nil} shows that the [000]1/2$^+$ $s$ orbit
is the lowest $\la$ s.p.~energy level
with a spherical density distribution concentrated at the center,
as seen in Fig.~\ref{f:orbit}.
As two hyperons can occupy this level,
and their mutual interaction is small,
the $\bav$ values of double-$\la$ hypernuclei
are very close to those of single-$\la$ hypernuclei.

Regarding the three negative-parity $p$ states,
Fig.~\ref{f:nil} shows that
the [101]1/2$^-$ and [101]3/2$^-$ orbits are degenerate
as the spin-orbit interaction is neglected in the $\la\la$ channel.
Their s.p.~energies are lower than those of the [110]1/2$^-$ orbit
on the oblate side, but higher on the prolate side.
Therefore a partial filling of the $p$ states
(4$\la$ and 6$\la$ hypernuclei)
allows to lower the total energy by increasing the magnitude of the deformation,
whereas a complete filling (8$\la$ hypernuclei) does not exhibit this feature.
As shown in Fig.~\ref{f:orbit},
the [110]1/2$^-$ orbit is prolate with zero density at $z=0$, while
the degenerate [101]1/2$^-$ and [101]3/2$^-$ orbits are both oblate
with zero densities at $r=0$.
When the 8 hyperons occupy fully the three $p$ orbits,
their density distribution becomes spherical.

\begin{figure*}[t]
\includegraphics[width=180mm]{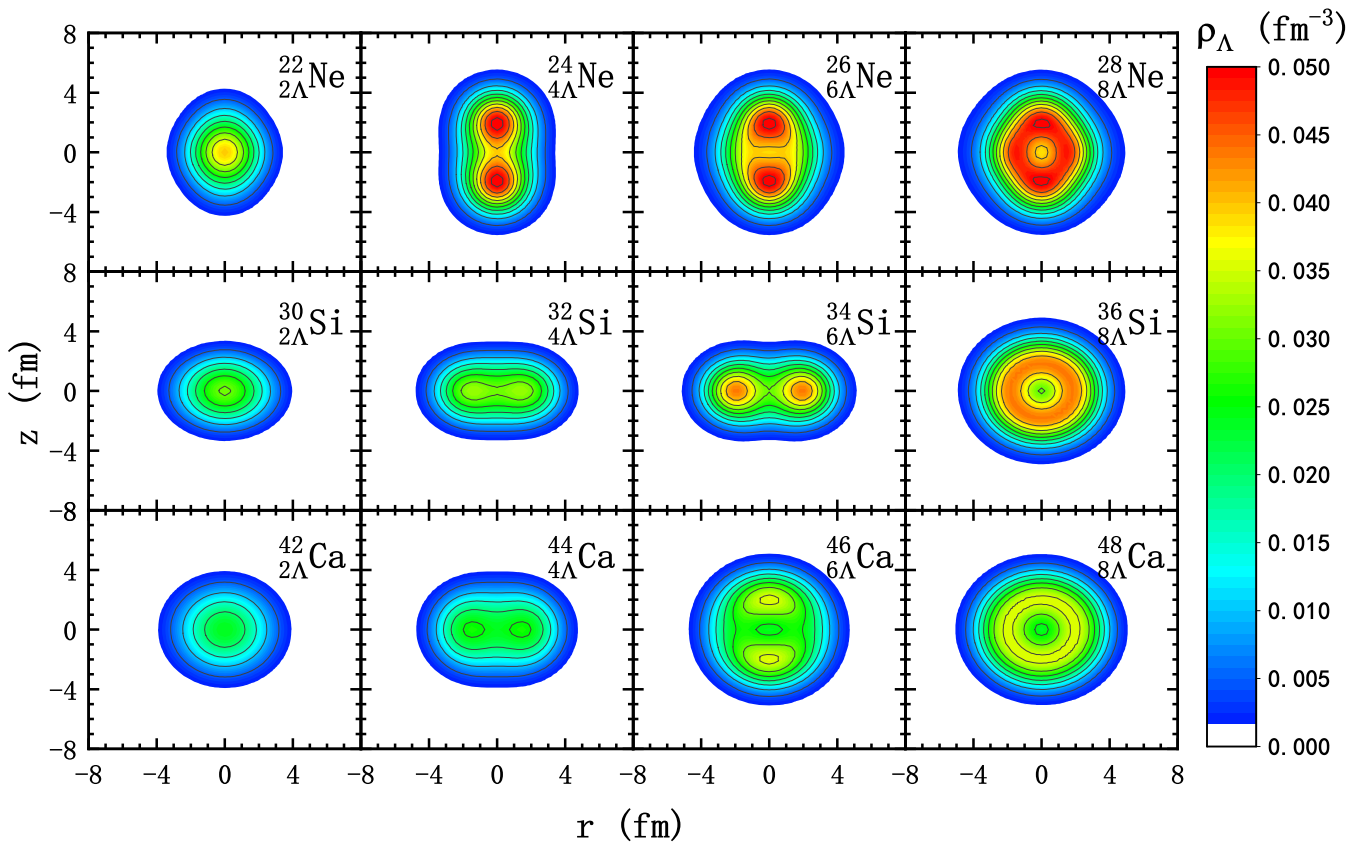}
\vskip-2mm
\caption{
Density distribution of hyperons in the $(r,z)$ plane
in Ne, Si, and Ca hyperisotopes.
The $z$ axis is the symmetry axis.}
\label{f:den}
\end{figure*}

A more detailed visualization of the effects of hyperons on the deformation
of hyperisotopes is given in Fig.~\ref{f:den},
which shows the $\la$ density distribution in the $(r,z)$ plane.
We choose the prolate Ne, the oblate Si, and the spherical Ca hyperisotopes
as examples.
It can be seen that the density distribution of double-$\la$ hypernuclei
changes in accordance with the deformation of the core nuclei,
since the additional double-$\la$ occupy the spherical [000]1/2$^+$ orbital.

When 4 $\la$ hyperons are filled in,
the shape of the first $p$ orbit occupied by the hyperons
is the same as that of the core nuclei.
For example,
the hyperons of $^{24}_{4\la}$Ne with prolate deformation first fill into
the [110]1/2$^-$ orbit, which is also prolate,
and then gradually fill into the [101]1/2$^-$ and [101]3/2$^-$ orbits,
which are oblate.
Thus the deformation of $^{24}_{4\la}$Ne reaches the largest value
due to the maximal distribution of 4 $\la$ hyperons in the prolate orbit.
When the hyperons begin to fill into the oblate orbits,
a reduction of the deformation occurs in $^{26}_{6\la}$Ne.
Finally, the spherical distribution of 8$\la$'s renders also the core nucleus
more spherical.

The hyperons in $^{28}$Si hyperisotopes with oblate core nucleus first fill
into the degenerate oblate orbits [101]1/2$^-$ and [101]3/2$^-$.
Therefore, the deformation increase can last up to 6$\la$ hypernuclei,
when the deformation reaches the maximum.
Then hyperons will fill into the prolate [110]1/2$^-$ orbit,
and cause a reduction of the deformation.
This also explains the different trends of deformation of prolate and oblate
hyperisotopes with the increasing hyperon number as shown in Fig.~\ref{f:eng}.

However, for spherical core nuclei, such as $^{16}$O and $^{40}$Ca,
the 4$\la$ hypernuclei have no preference for oblate or prolate orbits;
therefore their deformation trends have the characteristics
of both oblate and prolate hypernuclei.
As a consequence, they show a more or less soft potential energy surface
around the spherical shape,
in particular in the $^{16}$O hypernuclei as shown in Fig.~\ref{f:eng}.

\section{Summary}

In summary, we study
within the deformed SHF formalism
the impurity effects of $\la$ hyperons on the
deformation of even-even nuclei ranging from $^8$Be to $^{40}$Ca,
employing an effective $\la N$ interaction that reproduces well
the experimental binding energies of single-$\la$ hypernuclei.

The effects of $\la$ hyperons on the deformation of the core nuclei
are studied in detail.
Those deformations are generally reduced by adding 2 or 8 $\la$ hyperons,
while they are enhanced by adding 4 or 6.
These behaviors are interpreted in a microscopic manner
by analyzing the $\la$ s.p.~orbits
and the density distributions for the occupied ones.
It is demonstrated that the order of filling the hyperons into the $p$ orbits
is determined by the shape of the core nucleus.
When the core nucleus is oblate,
the hyperons are filled first into the degenerate oblate
[101]1/2$^-$ and [101]3/2$^-$ orbits,
so that the deformation of the core nucleus increases to the oblate side,
and can reach a maximum when 6 hyperons are added.
When the core nucleus is prolate,
the hyperons are filled first into the prolate [110]1/2$^-$ orbit,
which leads to an increase of the prolate deformation of the core nucleus,
and the deformation reaches a maximum when 4 hyperons are added.
When the core nucleus is spherical,
hyperons have no preferences for the shape of the $p$ orbits and result in
a soft potential energy surface.

Future experimental data of multi-$\la$ hypernuclei are necessary
to examine these effects and refine the assumptions and ingredients
of the model calculations presented here.

\bigskip
\section*{Acknowledgements}

This work was supported by the
National Natural Science Foundation of China under Grant Nos.~11775081, 1217051034 and 11905165.

\newcommand{\epja}{EPJA\ }
\bibliography{ref}

\end{document}